\documentclass[aps,prl,twocolumn,amsfonts,showpacs,superscriptaddress]{revtex4}
\usepackage[dvips]{graphicx}
\usepackage{bm}% bold math

\begin{document}
\title{Fluctuation Theorem in a Quantum-Dot Aharonov-Bohm Interferometer
}
\author{Yasuhiro Utsumi}
\address{Institute for Solid State Physics, University of Tokyo, Kashiwa, 
Chiba 277-8581, Japan}
\author{Keiji Saito}
\affiliation{
Graduate School of Science, University of Tokyo, Tokyo 113-0033, Japan}
\affiliation{
CREST, Japan Science and Technology (JST), Saitama 332-0012, Japan}

\begin{abstract}
In the present study, we investigate the full counting statistics in a two-terminal Aharonov-Bohm interferometer 
embedded with an interacting quantum dot.
We introduce a novel saddle-point solution for a cumulant-generating function, which satisfies the fluctuation theorem and accounts for the interaction in the mean-field level approximation. 
Nonlinear transport coefficients satisfy universal relations imposed by microscopic reversibility, though the scattering matrix itself is not reversible. 
The skewness can be finite even in equilibrium, owing to the interaction and is proportional to the asymmetric component of nonlinear conductance. 
\end{abstract}

\date{\today}
\maketitle

\newcommand{\mat}[1]{\mbox{\boldmath$#1$}}
\newcommand{\mtau}{\mbox{\boldmath$\tau$}}
\def\etal{et al.}

%{\it Introduction.}--
{\em Microscopic reversibility} is a key ingredient in deriving the Onsager relation and has played a fundamental role in establishing the linear response theory~\cite{Onsager}.
Recently, microscopic reversibility was used to develop a new relationship that would be valid beyond the linear response regime. 
This is now known as the {\em fluctuation theorem} (FT) \cite{Evans}. 
The FT relates probabilities between positive and negative entropy productions; provides a precise statement for the second law of thermodynamics; 
and remarkably, reproduces the linear response theory, 
the Kubo formula and the Onsager relation~\cite{Evans}.

In the last few years, {\em full counting statistics} (FCS) has been recognized as a suitable framework for the FT in quantum transport~\cite{Andrieux,Tobiska,Saito1,Saito,Foerster,Sanchez}. 
FCS provides a comprehensive statistical properties for charge transport far from equilibrium~\cite{Levitov,Noise,Utsumi1}. 
It addresses the probability distribution $P(q)$ of the charge $q$ which is transmitted during time $\tau$, and its cumulant generating function (CGF): 
${\cal F}(\lambda) = \lim_{\tau \to \infty} \ln {\cal Z}(\lambda)/\tau$, 
where
${\cal Z}(\lambda)=\sum_q P(q) \, {\rm e}^{i q \lambda}$
and $\lambda$ is called the counting field~\cite{Levitov}. 
Recently, the FT was generalized to the quantum transport regime in the presence of interaction and a magnetic field $B$~\cite{Saito}. 
For two-terminal systems, the FT is 
%----------------------------------------------------------
\begin{eqnarray}
{\cal F}({\lambda} ;B) 
\! &=& \!
{\cal F}( - {\lambda} + i {\cal A} ;-B) 
\, , 
\label{eqn:ft}
\\
P(q;B)
\! &=& \!
P(-q;-B) 
\, 
{\rm e}^{q {\cal A} }
\, , 
\label{eqn:ftp}
\end{eqnarray}
%----------------------------------------------------------
where ${\cal A}$ is the affinity ${\cal A} \!=\! V/T$, the ratio between 
voltage $V$ and temperature $T$ ($e \!=\! \hbar \!=\! k_B \!=\! 1$). 
One important consequence from (\ref{eqn:ft}) is the universal relations among transport coefficients~\cite{Saito}.
The transport coefficient $L$ is obtained by expanding the current cumulant with respect to ${\cal A}$: 
%----------------------------------------------------------
\begin{eqnarray}
\langle \! \langle
I^n
\rangle \! \rangle
=
\left.
{\partial^n {\cal F} (\lambda ;B) \over \partial (i \lambda)^n } 
\right|_{\lambda = 0}
=
\sum_{m=0}^\infty L^n_m(B) \, { {\cal A}^m \over m!}
\, ,
\nonumber
\end{eqnarray}
%----------------------------------------------------------
where $I \!=\! q/\tau$. 
The FT (\ref{eqn:ft}) leads to the Kubo formula 
$L^1_1 \!=\! L^2_0 /2$ 
and the Onsager relation 
$L^1_{1 , -} \!=\! 0$, 
where 
$L^n_{m , \pm} \!= \! L^n_m(B)  \pm  L^n_m(-B)$
is the symmetrized/antisymmetrized transport coefficient. 
Furthermore, nontrivial relations among higher-order coefficients are obtained~\cite{Saito}: 
%----------------------------------------------------------
\begin{eqnarray}
L^{1}_{2,-}\! = {1\over 3} L^{2}_{1,-}\! = {1\over 6} L^{3}_{0,-} \, ,
\;\;
L^{1}_{2,+} = L^{2}_{1,+} \, ,
\;\;
L^{3}_{0,+} = 0 \, . 
\label{eqn:relations}
\end{eqnarray}
%----------------------------------------------------------
This is significant in that the skewness $L^3_{0 , -}$ can be finite even in equilibrium and proportional to the asymmetric component of nonlinear conductance $L^1_{2 , -}$ as well as the linear response of noise $L^2_{1 , -}$. 

\begin{figure}[ht]
\includegraphics[width=.85 \columnwidth]{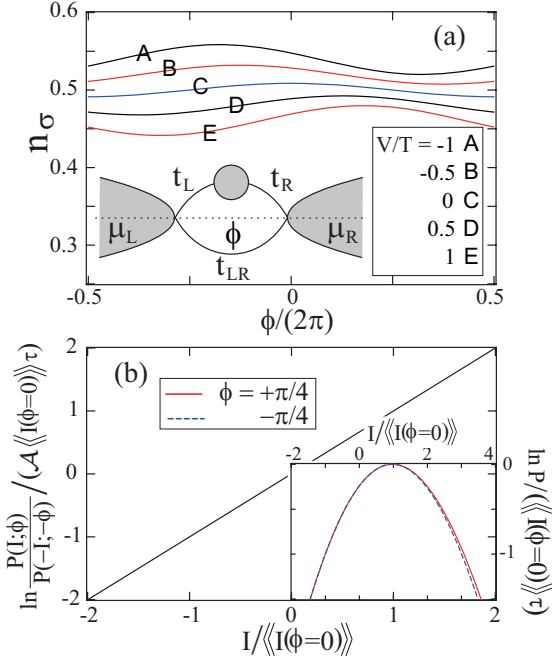}
\caption{
(a)
Aharonov-Bohm (AB) phase dependent nonequilibrium charge accumulation. 
(inset) Quantum-dot AB Interferometer. 
Mirror symmetry along the horizontal axis (dotted line) is absent. 
(b)
Demonstration of the fluctuation theorem (\ref{eqn:ftp}). 
The inset shows probability distributions for positive and negative magnetic fields. 
Parameters: 
$\Gamma_L/\Gamma \!=\! 0.25$, 
$\Gamma_R/\Gamma \!=\! 0.75$, 
$t_{\rm ref} \!=\! 0.25$, 
$\epsilon_D \!=\! 0$, 
and 
$U \!=\! T \!=\! V \!=\! \Gamma$.
}
\label{fig:figure1}
\end{figure}

However, there are some controversies regarding the validity of Eq.~(\ref{eqn:ft})~\cite{Foerster}. 
When interacting mesoscopic conductors possess no mirror symmetry, the nonequilibrium charge accumulation inside the conductor is not symmetric in the magnetic field~\cite{Leturcq}. 
Then, the potential landscape generated by the nonequilibrium charge accumulation is not symmetric either. 
This implies that the $S$-matrix is not reversible with respect to a magnetic field 
$S_{LR}(B) \neq S_{RL}(-B)$, which generates the magnetic field asymmetric component of nonlinear conductance and `violates' the Onsager relation~\cite{Buttiker,Spivak,Foerster,Leturcq,Rikken}. 
In this respect, the FT (\ref{eqn:ft}) is counterintuitive. 
In fact, in Ref.~\cite{Foerster}, the Mach-Zehnder interferometer is suggested as a counter example of the FT (\ref{eqn:ft}). 
Hence, it is necessary to give examples preserving the FT~(\ref{eqn:ft}). 
To this end, we consider a two-terminal Aharonov-Bohm (AB) interferometer embedded with a quantum dot (QD) [inset in Fig.~\ref{fig:figure1} (a)]~\cite{Koenig}. We introduce a novel saddle-point solution of CGF, which realizes the FT~(\ref{eqn:ft}) and the lack of reversibility in the $S$-matrix simultaneously. 
It is achieved by introducing the 'counting field of the dot charge' in addition to the dot potential, which are functions of $\lambda$. 
The solution accounts for nonequilibrium charge accumulation and current fluctuations in the Hartree-level approximation. 
We will also calculate Eq.~(\ref{eqn:relations}) explicitly and show that the equilibrium skewness is a consequence of Coulomb interaction.

{\it CGF of QD AB interferometer.}--
The system consists of left (L) and right (R) leads, two arms and a QD.
Electrons can travel through the QD and the lower reference arm [inset in Fig.~\ref{fig:figure1} (a)]. 
The total Hamiltonian is 
%----------------------------------------------------------
\begin{equation}
H   = \sum_{r=L,R} H_r + H_D + H_{\rm T} + H_{\rm ref} , 
\end{equation}
%----------------------------------------------------------
where the on-site Coulomb interaction $U$ in the QD is accounted for by 
$
H_D =
\sum_{\sigma = \uparrow , \downarrow} 
\epsilon_D 
d^{\dagger}_{\sigma} d_{\sigma} 
+
U d^{\dagger}_{\uparrow} d_{\uparrow} 
d^{\dagger}_{\downarrow} d_{\downarrow}
$. 
%----------------------------------------------------------
The operator $d_{\sigma}$ annihilates an electron with spin $\sigma$. 
The leads are modeled by 
$
H_r
=
\sum_{k \sigma}
\varepsilon_{r k \sigma} 
\, a^{\dagger}_{r k \sigma} \, a_{r k \sigma}
$, 
where $a_{r k \sigma}$ annihilates electrons in the lead $r$ with spin $\sigma$ and wave vector $k$. 
The tunneling and the reference arm are described as
$
H_{\rm T}
=
\sum_{r k \sigma}
t_r \, 
d_{\sigma}^\dagger
a_{r k \sigma} 
+{\rm H.c.}
$
and 
$
H_{\rm ref}
=
\sum_{k k' \sigma}
t_{LR} \, 
{\rm e}^{i \phi}
\, 
a_{R k \sigma}^\dagger
a_{L k' \sigma} 
+{\rm H.c.} \,
$. 
%---------------------------------------------------------- 
The magnetic field $B$ pierces through the ring, and the electrons acquire the AB phase $\phi$, which satisfies $\phi(B) \!=\! -\phi(-B)$. 
The initial density matrices at both leads are assumed to have an equilibrium distribution with the chemical potential $\mu_L = V/2$ and $\mu_R = -V/2$.
%----------------------------------------------------------
%----------------------------------------------------------

We confine ourselves to high temperature and treat the interaction in a mean-field level approximation~\cite{Anderson}. 
In order to find a proper saddle-point solution out of equilibrium, we employ the real-time path integral approach~\cite{Kamenev}. 
The characteristic function, which is the partition function in the Keldysh formalism, reads 
%----------------------------------------------------------
$
{\cal Z}
\! = \!
\int 
{\cal D}[a_{rk\sigma}^*,d_{\sigma}^*,a_{rk\sigma},d_{\sigma}]
\exp \left( i \int_C dt \, {\cal L}(t) \right)
$, 
%----------------------------------------------------------
where $C$ is the closed time-path. The Lagrangian is given by
%----------------------------------------------------------
\begin{eqnarray}
{\cal L}
\! &=& \!
\sum_{r k \sigma}
\, a_{r k \sigma}^* 
(i \partial_t-
\varepsilon_{r k \sigma} 
)
\, a_{r k \sigma}
\!+\!
\sum_\sigma
d_{\sigma}^* \, 
(i \partial_t-\epsilon_D)
d_{\sigma}
\nonumber \\
& &
-
\sum_{r k \sigma}
(t_r \, {\rm e}^{i \varphi_r} 
d_{\sigma}^* a_{r k \sigma}+c.c.)
-
U
d_{\uparrow}^* 
d_{\uparrow}
d_{\downarrow}^* 
d_{\downarrow}
\nonumber \\
\nonumber \\
& &
-
\sum_{k k' \sigma}
(t_{LR} \, {\rm e}^{-i \varphi_R+i \varphi_L+i \phi}
a_{R k \sigma}^* 
a_{L k \sigma}
+c.c.) \, , 
\nonumber
\end{eqnarray}
%----------------------------------------------------------
where the phase on the upper/lower branch of the closed time-path 
$\varphi_{r \pm}$ is related to the counting field as 
$\varphi_{r \pm} \!=\! \pm\lambda_r/2$. 
We introduce the auxiliary dot-potential $v_\sigma$ via the Stratonovich-Hubbard transformation: 
$
U
d_{\uparrow}^* 
d_{\uparrow}
d_{\downarrow}^* 
d_{\downarrow}
\rightarrow
\sum_\sigma
v_\sigma 
d_{\sigma}^* 
d_{\sigma}
\!-\!
v_\uparrow
v_\downarrow 
/U
$~\cite{Hamann}: 
%----------------------------------------------------------
\begin{eqnarray}
{\cal Z}
=
\int \! 
{\cal D}
[v_\sigma]
\,
{\cal Z}_0(v_\sigma)
\,
\exp
\left(
\frac{i}{U} \!\!
\int_C \!\! d t \, v_\uparrow(t) v_\downarrow(t)
\right)
\, ,
\nonumber
\end{eqnarray}
%----------------------------------------------------------
where ${\cal Z}_0(v_\sigma)$ is the Keldysh partition function for the noninteracting case $U\!=\!0$ with a shift in the QD level 
$\epsilon_D \! \rightarrow \! \epsilon_D \!+\! v_{\sigma}(t)$ for spin $\sigma$. 
Although we limit ourselves to the time-independent stationary solution in the non-magnetic phase~\cite{Anderson}, we allow different dot-potentials for upper and lower branches of $C$~\cite{Kindermann}: 
%----------------------------------------------------------
\begin{eqnarray}
v_{\sigma \, \pm}(t) = v_{\pm}=v_{c} \pm i v_{q}/2 \, .
\end{eqnarray}
%----------------------------------------------------------
The classical component $v_c$ is the dot potential generated by accumulation of charges with opposite spin. 
The quantum component $v_q$ plays the role of the counting field for charge in QD~\cite{Pilgram,Utsumi}. 
After a number of calculations, the CGF is represented by the $S$-matrix: 
%----------------------------------------------------------
\begin{eqnarray}
{\cal F}
\! &=& \! 
{\cal F}_0
-
\frac{2}{U} v_{c} v_{q}, 
\;\;\;\;
{\cal F}_0
=
\frac{1}{\pi}
\int \! 
d \omega \, 
{\rm Tr}
\ln
(
{\bf 1}-\tilde{f} \, {\bf K}
) \, ,
~~
\label{eqn:saddlecgf}
\\
{\bf K}
\! &=&\!
{\bf 1}
\! - \!
e^{i {\bf \tilde{\lambda}}} 
{\cal S}^\dagger(v_{-})
e^{-i {\bf \tilde{\lambda}}} 
{\cal S}(v_{+}),
\;\;
{\cal S}
\!=\!
\left(
\!
\begin{array}{cc}
S_{LL} & S_{LR} \\
S_{RL} & S_{RR}
\end{array}
\!
\right) \! ,
\label{eqn:s}
\end{eqnarray}
%----------------------------------------------------------
where 
${\bf 1}$ is a unit matrix, and 
$\tilde{\lambda} \!=\! {\rm diag}(\lambda,0)$
with 
$\lambda \!=\! \lambda_L \!-\! \lambda_R$. 
$\tilde{f} \!=\! {\rm diag}(f_L,f_R)$
consists of the Fermi distribution function
$f_r(\omega) \! = \! 1/ \{ \exp[(\omega - \mu_r)/T] + 1 \}$. 
When the potential $v$ is independent of the magnetic field, 
the $S$-matrix is reversible $S_{rr'}(v;B)=S_{r'r}(v;-B)$ 
%----------------------------------------------------------
\begin{eqnarray}
S_{rr}(v) 
\!\! &=& \!\!
1- 
\frac{
i \Gamma_r
+
t_{\rm ref} 
\sqrt{\Gamma_L \Gamma_R} \cos \phi
-
t_{\rm ref}^2 \, \epsilon(v)/2
}{\Delta(v)}, 
\nonumber
\\
S_{RL}(v) 
\!\! &=& \!\!
\left(
{\rm e}^{i \phi} t_{\rm ref} \, \epsilon(v) - \sqrt{\Gamma_L \Gamma_R}
\right)
/\Delta(v),
\nonumber 
\\
\Delta(v)
\!\! &=& \!\!
\frac{t_{\rm ref} \sqrt{\Gamma_{\! L} \Gamma_{\! R}} \, \cos \phi}
{2}
-
\biggl( \! 1 + \frac{{t_{\rm ref}}^2}{4} \! \biggl) \epsilon(v)
+
i \, \frac{\Gamma}{2},
\nonumber
\end{eqnarray}
%----------------------------------------------------------
where 
$\epsilon(v) \!=\! \epsilon_D \!+\! v \!-\! \omega$. 
The tunnel coupling 
$\Gamma \!=\! \Gamma_{\! L} \!+\! \Gamma_{\! R}$ 
is written with the DOS of the lead as 
$\Gamma_r \!=\! 2 \pi \, t_r^2 \, \rho_r$. 
Hopping through the reference arm is characterized by 
$t_{\rm ref} \!=\! 2 \pi \, t_{LR} \sqrt{\rho_L \rho_R}$. 
It appears that Eq.~(\ref{eqn:s}) is the CGF for the joint probability distribution of current and charge~\cite{Pilgram,Utsumi}. 
However, the 'charge counting field' $v_q$ as a function of $\lambda$ is now determined by coupled saddle-point equations: 
%----------------------------------------------------------
\begin{eqnarray}
v_{c}
=
(U/2)
\, 
\partial {\cal F}_0/
\partial v_{q}
\, , 
\;\;
v_{q}
=
(U/2)
\, 
\partial {\cal F}_0/
\partial v_{c}. 
\label{eqn:saddle}
\end{eqnarray}
%----------------------------------------------------------

{\it Magnetic field asymmetry in nonlinear transport.}--
The saddle-point solution (\ref{eqn:saddle}) captures magnetic field asymmetry in the nonlinear transport regime~\cite{Buttiker,Foerster}. 
For $\lambda \!=\! 0$, Eq.~(\ref{eqn:saddle}) possesses a trivial solution: 
$v_q \!=\ 0$ and $v_c \!=\! v^*$ determined by the nonequilibrium Hartree equation, 
$v^* \!=\! U \int \! d \omega A_\sigma(\omega)/(2 \pi)$,
%----------------------------------------------------------
\begin{eqnarray}
A_\sigma
\!\! 
&=& 
\!\!\!
\sum_{r}
(
\Gamma_{\! r}
+
t_{\rm ref}^2
\Gamma_{\! \bar{r}}/4
)
[f_r(\omega)-1/2]
/|\Delta(v^*)|^2
\nonumber \\ 
& & 
+ \, t_{\rm ref} \sqrt{ \Gamma_{\! L} \Gamma_{\! R} } \sin \phi 
\, 
[f_L(\omega) \! - \! f_R(\omega)] 
/|\Delta(v^*)|^2,~~~
\label{eqn:nd}
\end{eqnarray}
%----------------------------------------------------------
where 
$\bar{r} \!=\! L/R$ for $r \!=\! R/L$. 
Figure~\ref{fig:figure1} (a) shows the magnetic field dependence of charge accumulation inside the QD, $n_\sigma =v^*/U \!+\! 1/2$. 
In equilibrium $V \!=\! 0$, $n_\sigma$ is an even function of the magnetic field. 
For $V \! \neq \! 0$, because the second line of Eq.~(\ref{eqn:nd}) is related to the lack of mirror symmetry, the charge accumulation becomes an uneven function of AB flux
$n_\sigma(\phi) \! \neq \! n_\sigma(-\phi)$. 

The average of the charge current is obtained by differentiating the CGF in terms of $\lambda$. 
%----------------------------------------------------------
\begin{equation}
\frac{d {\cal F}}{d (i \lambda)}
\!=\!
\frac{\partial {\cal F}_0}{\partial (i \lambda)}
+
\!\!
\sum_{\alpha=c,q}
\!\!
\left (
\frac{\partial {\cal F}_0}{\partial v_{\alpha}}
\frac{d v_{\alpha}}{d (i \lambda)}
-
2
\frac{
v_{\bar{\alpha}}
}{U}
\frac{d v_{\alpha}}{d (i \lambda)}
\right)
\!=\!
\frac{\partial {\cal F}_0}{\partial (i \lambda)} \, ,
\label{eqn:1st}
\end{equation}
%----------------------------------------------------------
where 
$\bar{\alpha} \!=\! c/q$ for $\alpha \!=\! q/c$. 
All contributions except ${\cal F}_0$ cancel because of the condition~(\ref{eqn:saddle}). 
Then, the Landauer formula with the transmission probablity ${\cal T}=|S_{LR}(v^*)|^2$ is obtained;  
%----------------------------------------------------------
\begin{eqnarray}
\langle \! \langle
I
\rangle \! \rangle
\! &=& \!
\left.
\frac{d {\cal F}_0}{d (i \lambda)}
\right |_{\lambda=0}
=
\frac{1}{\pi} \! \int \! d \omega
\, 
{\cal T}(\omega)
\, 
[f_{L}(\omega)-f_{R}(\omega)]
\, ,
\nonumber \\
{\cal T}(\omega)
\! &=& \!
\frac{
\Gamma_{\! L} \Gamma_{\! R} 
+
t_{\rm ref}^2 \, \epsilon(v^*)^2
-
2 \, t_{\rm ref} \, \epsilon(v^*)
\sqrt{ \Gamma_{\! L} \Gamma_{\! R} }
\cos \phi
}{|\Delta(v^*)|^2} 
\, .
\nonumber
\end{eqnarray}
%----------------------------------------------------------
In the absence of interaction $U\!=\!0$, and thus $v \!=\! 0$, the transmission probability is symmetric in the magnetic field~\cite{Koenig}. 
For finite $U$ and $V$, because of the accumulation of magnetic field-dependent nonequilibrium charges, the reversibility of $S$-matrix breaks down as 
$S_{LR}(v^*(B);B) \! \neq \! S_{RL}(v^*(-B);-B)$,
leading to magnetic field asymmetry in nonlinear conductance.

If we substitute $v_q \!=\! 0$ and $v_c \!=\! v^*$ in Eqs.~(\ref{eqn:saddlecgf}) and (\ref{eqn:s}), our CGF may be compatible with that in Ref.~\cite{Foerster} at the formal level. 
However, for our case, generally both $v_c$ and $v_q$ satisfying Eq.~(\ref{eqn:saddle}) depend on $\lambda$. 
Then the CGF (\ref{eqn:saddlecgf}) with Eq.~(\ref{eqn:saddle}), fulfills the FT~(\ref{eqn:ft}), since if we consider $v_\pm$ as variables, 
$
{\cal F}_0(\lambda,v_\pm;B)
\!=\!
{\cal F}_0(-\lambda+i {\cal A},v_\pm;-B)
$
is satisfied for any $v_\pm$. 
Figure~\ref{fig:figure1} (b) demonstrates the FT~(\ref{eqn:ftp}), though probability distributions for positive and negative magnetic fields are different [inset of Fig.~\ref{fig:figure1} (b)]. 
Therefore, the magnetic field asymmetry does not necessarily contradict the FT.

{\it Nonequilibrium noise.}--
In the presence of interaction, current and charge fluctuations couple in a nontrivial manner, which means that we must account for $v_q$ carefully. 
Let us consider the derivative of Eq.~(\ref{eqn:saddle}) with respect to the counting field:
%----------------------------------------------------------
\begin{eqnarray}
\frac{d v_{\bar{\alpha}}}{d (i \lambda)}
\! &=& \!
\frac{U}{2} 
\frac{\partial^2 {\cal F}_0}{\partial v_{\alpha} \, \partial  (i \lambda)}
+
\frac{U}{2} \!
\sum_{\alpha'=c,q}
\frac{\partial^2 {\cal F}_0}{\partial v_{\alpha} \, \partial v_{\alpha'}}
\frac{d v_{\alpha'}}{d (i \lambda)}
\, ,~~~
\label{eqn:vertex_}
\\
\! &=& \!
\sum_{\alpha'}
\frac{U_{\alpha \alpha'}}{2}
\frac{\partial^2 {\cal F}_0}{\partial v_{\alpha'} \, \partial (i \lambda)}
\, .
\label{eqn:vertex}
\end{eqnarray}
%----------------------------------------------------------
For $\lambda=0$, 
which implies that $v_q=0$ and $v_c=v^*$, the four components are
$U_{cc}|_{\lambda=0} \!=\! \tilde{U} S_{NN}$, 
$U_{cq}|_{\lambda=0} \!=\! U_{cq}|_{\lambda=0} \!=\! \tilde{U}$
and
$U_{qq}|_{\lambda=0} \!=\! 0$. 
Coulomb interaction is screened 
$\tilde{U} \!=\! U/(1 - U \chi_{NN})$, 
because the right-hand side of~(\ref{eqn:vertex_}) contains the derivative of $v_\alpha$ itself. 
The bare density-density response function $\chi_{NN}$ 
and the density-density correlation function (charge noise) $S_{NN}$ 
are given by 
$
\chi_{NN} 
\! = \!
\partial^2 {\cal F}_0/\partial v_c \partial v_q
|_{\lambda=0}/2
\! = \!
\partial \, n_\sigma/\partial \epsilon_D 
$ 
and
%----------------------------------------------------------
$
S_{NN}
\! = \!
\partial^2 {\cal F}_0/\partial v_q^2 
|_{\lambda=0}/2
$. 
%----------------------------------------------------------
%
Then, with the help of Eq.~(\ref{eqn:vertex}), the derivative of Eq.~(\ref{eqn:1st}) 
and the full form of the nonequilibrium current noise read as follows: 
%----------------------------------------------------------
\begin{eqnarray}
\frac{d^2 {\cal F}}{d (i \lambda)^2}
\!\! &=& \!\!
\frac{\partial^2 {\cal F}_0}{\partial (i \lambda)^2}
+
\sum_{\alpha, \alpha'}
\frac{\partial^2 {\cal F}_0}{\partial (i \lambda) \, \partial v_{\alpha}}
\,
\frac{U_{\alpha \alpha'}}{2}
\,
\frac{\partial^2 {\cal F}_0}{\partial (i \lambda) \, \partial v_{\alpha'}},
~~~~
\label{eqn:2nd}
\\
\langle \! \langle I^2 \rangle \! \rangle
\!\! &=& \!\!
2 \, (
S_{II}
+
2 \, S_{IN} \, \tilde{U} \, \chi_{IN}
+
{\chi_{IN}}^2 S_{NN} \, \tilde{U}^2
)
\, ,
\label{eqn:sjj}
\end{eqnarray}
%----------------------------------------------------------
where the bare current-density response and the current-density correlation functions are 
$
\chi_{IN} \!=\! (\partial \, \langle \! \langle I \rangle \! \rangle/\partial \epsilon_D)/2
$
and
$
S_{IN} \! = \! \partial^2 {\cal F}_0/\partial (i \lambda) \partial v_{q} \, |_{\lambda=0}
/2
$. 
The current-current correlation, 
%----------------------------------------------------------
\begin{eqnarray}
S_{II}
\! &=& \!
\frac{1}{2}
\frac{\partial^2 {\cal F}_0}{\partial (i \lambda)^2}
\biggl|_{\lambda=0}
=
\frac{1}{2 \pi}
\int \!\! d \omega \, 
{\cal T}(\omega) \, 
[f_L(\omega)+f_R(\omega)
\nonumber \\
& &
-2 f_L(\omega) f_R(\omega)]
-
{\cal T}(\omega)^2
[f_L(\omega) - f_R(\omega)]^2
\, ,
\nonumber
\end{eqnarray}
%----------------------------------------------------------
is given by the quantum noise formula for noninteracting systems~\cite{Lesovik} with the self-consistent potential $v^*$. 
The second and third terms of Eq.~(\ref{eqn:sjj}) are the result of interaction out of equilibrium, since in equilibrium, the average current vanishes and consequently, $\chi_{IN} \!=\! 0$. 
In the absence of the reference arm $t_{\rm ref}\!=\!0$, Eq.~(\ref{eqn:sjj}) reproduces the theory of the noise for the nonequilibrium Anderson model in the Hartree-level approximation~\cite{Hershfield}. 
For this, the counting filed of QD charge $v_q$ is crucial.

\begin{figure}[ht]
\includegraphics[width=.75 \columnwidth]{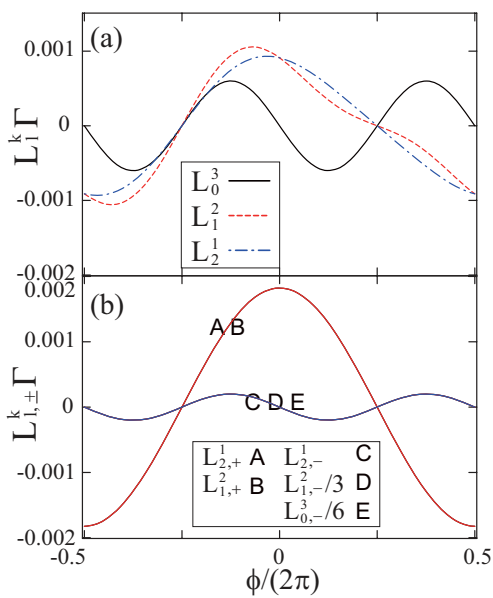}
\caption{
(a) 
Aharonov-Bohm flux dependent third-order nonlinear transport coefficients and (b) the extension of Onsager's theorem. 
The parameters are the same as those in Fig.~\ref{fig:figure1}. 
}
\label{fig:figure2}
\end{figure}

{\it Nonlinear transport coefficients.}--
Now we come to the relations among the third-order transport coefficients~(\ref{eqn:relations}). 
First, the bare parts vanish: 
${\partial_{i \lambda}}^{\!\!\! 3-n} \partial_{\cal A}^{\, n} 
{\cal F}_0(0,B)|_{{\cal A}=0} 
\!=\! 0$ ($n=0,1,2,3$).
Then, the skewness, following the derivative of Eq.~(\ref{eqn:2nd}) in terms of $\lambda$ reads 
%----------------------------------------------------------
\begin{eqnarray}
L^3_0
=
6 \, 
\tilde{U}^{eq.} 
S_{IN}^{\, eq.} \, \chi_{II,N}^{eq.}
\, ,
\;\;\;\;
\chi_{II,N}
=
\partial \, S_{II}/\partial \epsilon_D \, ,
\label{eqn:skewness}
\end{eqnarray}
%----------------------------------------------------------
where $\chi_{II,N}$ is the linear response of the noise. 
%~\cite{Gabelli}. 
The superscript $eq.$ specifies that ${\cal A}$ is fixed at 0. 
Equation (\ref{eqn:skewness}) reveals that equilibrium skewness is caused by the interaction. 
The other transport coefficients are calculated in the same manner. 
%----------------------------------------------------------
\begin{eqnarray}
L^2_1
\! &=& \!
2 \, \tilde{U}^{eq.} \, \chi_{NI}^{eq.} \, \chi_{II,N}^{eq.}
+
4 \, \tilde{U}^{eq.} \, S_{IN}^{\, eq.} \, \chi_{I,IN}^{eq.} \, ,
\\
L^1_2
\! &=& \!
4 \, \tilde{U}^{eq.} \, \chi_{NI}^{eq.} \, \chi_{I,IN}^{eq.}
\, ,
\end{eqnarray}
%----------------------------------------------------------
where 
%----------------------------------------------------------
$\chi_{NI} \!=\! {\partial \, n_\sigma}/{\partial {\cal A}}$
and 
$\chi_{I,IN} \!=\! {\partial \, \chi_{IN}}/{\partial {\cal A}}$.
%----------------------------------------------------------
%
Figure~\ref{fig:figure2} (a) shows the AB flux dependence of third-order nonlinear transport coefficients. 
We observe finite skewness for $\phi \neq 0$ [panel (a)]. 
It appears that the coefficients behave independently. 
However, as shown in panel (b), an extension of the Onsager relation~(\ref{eqn:relations}) is satisfied perfectly. 

%----------------------------------------------------------
%----------------------------------------------------------

We note that our results can be obtained using the Hartree approximation based on the nonequilibrium self-consistent $\Phi$-derivable approximation~\cite{Ivanov,Baym}. 
In this scheme, the Keldysh generating function consists of an infinite number of closed diagrams, each of which satisfies the symmetry (\ref{eqn:ft}), as shown in Ref.~\cite{Saito}.

{\it Summary}--
We studied the full counting statistics of a quantum dot Aharonov-Bohm interferometer and have developed a novel Hartree approximation, which satisfies the fluctuation theorem and describes magnetic field asymmetry in the nonlinear transport. 
We have also shown that equilibrium skewness as well as the asymmetric component of nonlinear conductance are the result of Coulomb interaction. 
These satisfy the extension of Onsager relations (\ref{eqn:relations})~\cite{Saito}, which may be measured by the currently available experiments~\cite{Leturcq}. 

We thank T~Fujii, D.~S.~Golubev, and E.~Iyoda for valuable comments. 
This research was supported by Strategic International Cooperative Program JST.


\begin{thebibliography}{99}

\bibitem{Onsager}
L. Onsager, Phys. Rev. {\rm 37}, 405, (1931); H. B. G. Casimir, Rev. Mod. Phys. {\rm 17}, 343, (1945)

\bibitem{Evans}  
D. J. Evans, E G. D. Cohen, and G. P. Morriss, Phys. Rev. Lett. {\bf 71}, 2401 (1993); G. Gallavotti and E. G. D. Cohen, Phys. Rev. Lett. {\bf 74}, 2694 (1995); S. Yukawa, J. Phys. Soc. Jpn. {\bf 69}, 2363 (2000); H. Tasaki, cond-mat/0009244; T. Monnai and S. Tasaki, cond-mat/0308337; C. Jarzynski and D. K. W\'ojcik, Phys. Rev. Lett. {\bf 92}, 230602 (2004). 

\bibitem{Andrieux}
D. Andrieux and P. Gaspard, J. Stat. Mech. P01011 (2006); J. Stat. Mech. P02006 (2007); M. Esposito, U. Harbola, and S. Mukamel, Phys. Rev. B {\bf 75}, 155316 (2007). 

\bibitem{Saito1}
K. Saito and A. Dhar, Phys. Rev. Lett., {\bf 99}, 180601 (2007). 

\bibitem{Saito}
K. Saito and Y. Utsumi, arXiv:0709.4128; Phys. Rev. B {\bf 78}, 115429 (2008). 

\bibitem{Tobiska}
J. Tobiska and Yu. V. Nazarov, Phys. Rev. B {\bf 72}, 235328 (2005).

\bibitem{Foerster}
H. F\"orster and M. B\"uttiker, arXiv:0805.0362. 

\bibitem{Sanchez}
D.~S\'anchez, arXiv:0805.0788. 

\bibitem{Levitov} 
L. S. Levitov and G. B. Lesovik, JETP Lett. {\bf 58}, 230 (1993); L. S. Levitov, H.-W. Lee, and G. B. Lesovik, Journal of Mathematical Physics, {\bf 37}, 4845 (1996).

\bibitem{Noise} {\it Quantum Noise in Mesoscopic Physics}, Vol. 97 of {\it NATO Science Series II: Mathematics, Physics and Chemistry} edited by Yu. V. Nazarov (Kluwer Academic Publishers, Dordrecht/Boston/London, 2003).

\bibitem{Utsumi1} 
D.~Bagrets \etal in {\it Elements of Quantum Information}, eds. W. P. Schleich, H. Walther (Wiley-VCH, 2007); Y. Utsumi, D. S. Golubev, and G. Sch\"on, Phys. Rev. Lett. {\bf 96}, 086803 (2006). 

\bibitem{Leturcq}
R. Leturcq \etal, Phys. Rev. Lett. {\bf 96}, 126801 (2006); A. L\"{o}fgren \etal, Phys. Rev. Lett. {\bf 92} 046803 (2004); 
%C. A. Marlow, R. P. Taylor, M. Fairbanks, I. Shorubalko, and H. Linke, 
C. A. Marlow \etal, 
Phys. Rev. Lett. {\bf 96} 116801 (2006). 
%A. L\"ofgren \etal, Phys. Rev. B {\bf 73}, 235321 (2006). 


\bibitem{Buttiker} 
D. S\'{a}nchez and M. B\"uttiker, Phys. Rev. Lett. {\bf 93}, 106802 (2004); M. L. Polianski and M. B\"uttiker, Phys. Rev. Lett. {\bf 96}, 156804 (2006). 

\bibitem{Spivak} 
B. Spivak and A. Zyuzin, Phys. Rev. Lett. {\bf 93}, 226801 (2004). 
%E. Deyo, B. Spivak, and A. Zyuzin, Phys. Rev. B {\bf 74}, 104205 (2006). 


\bibitem{Rikken}
%G. L. J. A. Rikken, J. Folling, P. Wyder, 
G. L. J. A. Rikken \etal, 
Phys. Rev. Lett. {\bf 87}, 236602 (2001).

\bibitem{Koenig}
J. K\"onig and Y. Gefen, Phys. Rev. B {\bf 65}, 045316 (2002). 

\bibitem{Kamenev}
A. Kamenev in {\it Nanophysics: Coherence and Transport,} 
(Les Houches, Volume Session LXXXI) eds. H. Bouchiat \etal,
(Elsevier, Amsterdam, 2005); K.-C. Chou \etal, Phys. Rep. {\bf 118}, 1 (1985);
G. Sch\"on and A. D. Zaikin, Phys. Rep. {\bf 198}, 237 (1990). 

\bibitem{Hamann}
D. R. Hamann, Phys. Rev. Lett. {\bf 23}, 95 (1969).

\bibitem{Anderson}
P. W. Anderson, Phys. Rev. {\bf 124}, 41 (1961); A. Komnik and A. O. Gogolin, Phys. Rev. B {\bf 69}, 153102 (2004). 

\bibitem{Kindermann}
M. Kindermann, Yu. V. Nazarov, and C. W. J. Beenakker, Phys. Rev. Lett. {\bf 90}, 246805 (2003).

\bibitem{Pilgram} 
S. Pilgram, and M. B\"uttiker, Phys. Rev. B {\bf 67}, 235308 (2003). 

\bibitem{Utsumi}
Y. Utsumi, Phys. Rev. B {\bf 75}, 035333 (2007). 


\bibitem{Lesovik}
G.~B.~Lesovik, JETP Lett. {\bf 49}, 592 (1989).

%\bibitem{Gabelli} J.~Gabelli and B.~Reulet, Phys. Rev. Lett. {\bf 100}, 026601 (2008). 

\bibitem{Hershfield}
S. Hershfield, Phys. Rev. B {\bf 46}, 7061 (1992). 

\bibitem{Baym} 
G. Baym and L. P. Kadanoff, Phys. Rev. {\bf 124}, 287 (1961); G. Baym, Phys. Rev. {\bf 127}, 1391 (1962). 

\bibitem{Ivanov} 
Yu. B. Ivanov, J. Knoll, and D. N. Voskresensky, Nucl. Phys. A {\bf 657}, 413 (1999). 


\end{thebibliography}
\end{document}